# Transparent Conductive Oxides-based Architectures for the Electrical Modulation of the Optical Response: A Spectroscopic Ellipsometry Study

Running title: Optical properties of TCOs-based Systems

Running Authors: Maria Sygletou et al.

Maria Sygletou [1,a)], Francesco Bisio [2], Stefania Benedetti [3], Piero Torelli [4], Alessandro di Bona [3], Aleksandr Petrov [4] and Maurizio Canepa [1]

[1]OptMatLab, Dipartimento di Fisica, Università di Genova, via Dodecaneso 33, 16146 Genova, Italy
[2]CNR-SPIN, C.so Perrone 24, 16152 Genova, Italy
[3]CNR-Istituto Nanoscienze, via Campi 213/a 41125 Modena, Italy
[4]CNR-Istituto Officina dei Materiali, Laboratorio TASC in Area Science Park, S.S. 14 km 163.5, Basovizza, 34149 Trieste, Italy

[a)] Electronic mail: sygletou@fisica.unige.it

Transparent Conductive Oxides (TCOs) are a class of materials that combine high optical transparency with high electrical conductivity. This property makes them uniquely appealing as transparent-conductive electrodes in solar cells and interesting for optoelectronics and infrared-plasmonics applications. One of the new challenges that researchers and engineers are facing is merging optical and electrical control in a single device for developing next-generation photovoltaic, opto-electronic devices and energy-efficient solid-state lighting. In this work, we investigated the possible variations in the dielectric properties of aluminum-doped ZnO (AZO) upon gating, by means of Spectroscopic Ellipsometry (SE). We investigated the electrical-bias-dependent optical response of thin AZO films fabricated by magnetron sputtering, within a parallel-plane capacitor configuration. We address the possibility to control their optical and electric performances by applying bias, monitoring the effect of charge injection/depletion in the AZO layer by means of *in-operando* SE *vs* applied gate voltage.

## I. INTRODUCTION

The capability of chemically and physically engineering the electrical and optical response of solids represents both the driving force and the consequence of the tremendous technological development of electronics, optoelectronics and plasmonics within the past century. As a most relevant example, the process—known as doping—of introducing foreign impurities in suitable hosts has led, among others, to the development of TCOs, that are nowadays an irreplaceable component of solar cells and touch screens. TCOs are a unique class of materials characterized by visible-light optical transparency up to 80-90% combined with electrical resistivity as low as $10^{-4}$ Ohm·cm[1]. The amazing properties of TCOs are typically achieved adding suitable metal dopants to a wide-gap semiconducting oxide like ZnO, $TiO_2$ etc, and can be further tuned by changing the carrier concentration through application of external stimuli (e.g. electric fields, electron injection etc.) or even combining them with noble-metal nanoparticles[2–10] in order to control and improve their optical and electric properties.

At present, Indium-Tin Oxide ranks as the TCO with the best performances on record. However, the shortage of supply and increasing cost of In have driven the scientific interest towards different materials and combinations such as Aluminum-doped Zinc Oxide (AZO)[11]. Its plasma frequency falls in the deep infrared region and the material is cheap and easy to fabricate, e.g. by dc magnetron sputtering, atomic-layer and pulsed-laser deposition.

One of the new challenges that researchers are facing is merging optical and electrical control in a single device for developing next-generation photovoltaic, opto-electronic devices and energy-efficient solid-state lighting. The modification of the refractive index of optical materials by applying electric field has been a constant focus in the field of photonics. The carrier densities and the resulting refractive index changes upon gating are inherently small for conventional semiconductors (such as Si, GaAs etc.). In contrast, TCOs seem to be an interesting alternative, compared to conventional semiconductors, since it has already been reported that changes in carrier density of conductive oxides, with carrier densities in the $10^{21}$-$10^{22}$ cm$^{-3}$ range, yield strong variations – on the order of unity – in the refractive index by applying electric field[12].

Here, we propose the realization of a parallel plane capacitor system, with AZO layer as the top gate, tuning its optical properties by applying voltage. The system consists of the active TCO layer, a high permittivity insulator and a conductive-oxide substrate. SE (0.73–5.05 eV spectral range) is applied for the investigation of the dielectric properties of each individual layer of the proposed capacitor. It has already been used to determine the dielectric functions of ZnO films[13–16], while up to now only few studies have been reported to extract the optical constants of AZO films by this method[10,17]. In addition, the dielectric constants, extracted from SE, are used for modeling the expected optical response of the capacitor upon dc gating. Although the optical properties of these materials (Nb-STO, BTO, AZO) are relatively known, the study of them combined in one single device and the investigation of the optical and electrical response of such devices upon bias has not been achieved yet. This investigation could be the first step for the development of next generation optoelectronic devices with electrically-controlled properties. Our study for the moment is limited to the exploitation of the expected response of these systems upon bias that can be a guide for their future fabrication. These simulations represent a useful benchmark for the determination of the optimum configuration for monitoring any variation in the optical properties of the TCO upon gating. The electrical control of the electrical and optical properties of TCOs can lead to the fabrication of active TCO-based optical systems. This new class of materials could be very promising for optoelectronic applications and telecommunications.

## II. EXPERIMENTAL

The optical constants of each layer of the proposed capacitor were extracted from aptly-prepared reference samples. In particular, the optical properties of ZnO and AZO (4% Al-doped ZnO) were measured on films deposited by RF magnetron sputtering of ZnO (99.99%) and $Al_2O_3$ (99.99%) ceramic targets[18,19] on magnesium oxide (MgO) substrates. More details on the deposition conditions can be found in previous studies[18,20]. The insulating layer was a ceramic $BaTiO_3$ (BTO) film deposited by Molecular Beam Epitaxy (MBE) on a commercially available conductive crystal of 0.5% Nb-doped $SrTiO_3$ (NSTO)[21,22].

The BTO is a ferroelectric ceramic chosen for its large dc permittivity constant (300-800 at room temperature and up to several thousand at low temperature). This insulating material will determine then the capacitance of the whole TCO-based stack, allowing a large electron accumulation at the electrodes, not possible otherwise. Applying a voltage between the top and bottom-gate, a redistribution of carriers at both interfaces occurs. When a positive (negative) potential is applied to the electrodes, an accumulation (depletion) of carriers occurs at the AZO/BTO interface. This variation of carrier density leads to a change of the dielectric constant of AZO. A rough estimate of the gating-induced charge in the AZO can be obtained modelling the system as a parallel-plate capacitor. A voltage gate $V_g$ applied to the plates separated by an insulator of thickness $d$ and relative permittivity $\varepsilon_r$, induces on the plates a surface charge density $\sigma_e$ according to

$$V_g = \frac{\sigma_e d}{\varepsilon_r \varepsilon_0} \qquad (1)$$

where $\varepsilon_0$ is the vacuum permittivity. For our simulations, the level of Al concentration $4 \pm 0.5\%$ was chosen because it has already been reported that it is the optimum doping level for highly conductive and transparent AZO films[18]. Increasing doping above this value induces the appearance of states in the gap near the valence band maximum, which are related to the defects introduced in the oxide, that reduce the number of free carriers and probably also their mobility[20].

The target is to obtain electrically-functional AZO films as thin as 10 nm on top of BTO film of 100 nm thickness and relative permittivity as high as 500. Although these materials were alternatively combined in thin-film systems and devices[21,23–30], the stacking of the three oxides with the desired thicknesses, electronic properties and interface microstructure is a task for which no references in literature currently exist.

The optical characterization of the reference layers was performed by means of a J.A. Woollam M-2000 ellipsometer (0.73–5.05 eV, incidence angles of 50°, 60° and 70°). Benefitting from our experience on the investigation of the optical properties of other complex systems[31–33], we used the WVASE software (J.A. Woollam, Co.) for the SE modelling and simulations, allowing a thorough characterization of the optical constants $\varepsilon_1$ and $\varepsilon_2$ (complex refraction index), film thickness and roughness of the materials involved.

## III. RESULTS AND DISCUSSION

### A. Optical properties of the reference materials

The complex dielectric functions of the materials constituting the capacitor, were determined using SE in the 0.73–5.05 eV photon energy range. In Fig. 1, the ellipsometric angles, $\Psi$ and $\Delta$, of Nb:SrTiO$_3$ substrate (Fig. 1a, b), BaTiO$_3$ film on Nb:SrTiO$_3$ (Fig. 1d, e) as well as pure ZnO (Fig. 1g, h) and 4% Al-doped ZnO (Fig. 1j, k) films on MgO substrates, are shown.

For the parameterization of the optical properties of NSTO and ZnO, 2 PSEMI-M0 oscillators were employed, for BTO 3 PSEMI-M0 and a Lorentz oscillator while for AZO 2PSEMI-M0 and a Drude oscillator. Lorentz and PSEMI-M0 oscillators are suitable for the optical characterization of semi-conductive materials. In particular, Lorentz is the classic Lorentz oscillator and PSEMI-M0 is a parametrized semiconductor oscillator model for modelling crystalline semiconductors. The fit parameters of PSEMI-M0 are: A (amplitude), Br (broadening), E0 (oscillator energy), PR, WR, AR and O2R (polynomial control point parameters). In addition, the Drude function is a zero-resonance-energy Lorentz oscillator, used to represent free carriers, with fit parameters A (amplitude) and Br (broadening).

The spectra of Ψ and Δ for the reference samples at incident angles of 50°, 60° and 70°, are shown in Fig. 1. The thickness of BTO sample determined by normal fit is 115 nm, while the thickness, calculated by X-Ray Reflectivity-XRR is 120 nm (Fig. 1). For ZnO and AZO films the thickness calculated by SE is 300 nm and 330 nm, respectively, while the thickness, determined by profilometry, is approximately 300 nm in the case of ZnO and 340 nm in the case of AZO. The surface roughness of ZnO and AZO, according to Atomic Force Microscopy (AFM) measurements, is ~ 4 nm which is in a good agreement with the values deduced optically by SE (~6 nm) (Table I).

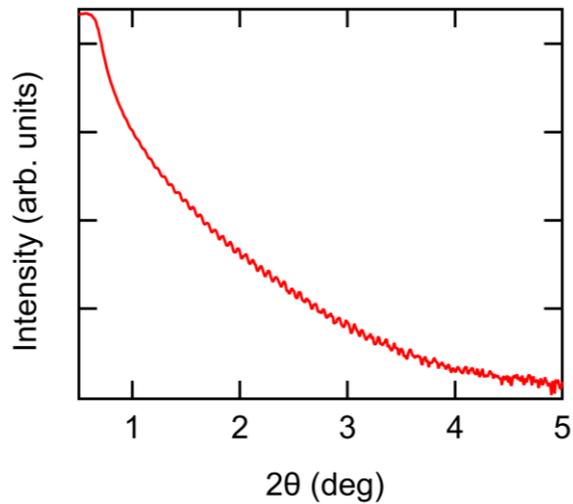

FIG. 1 XRR measurement of BTO film for the determination of the film thickness (y-axis in logarithmic scale).

Our fitting results of the real ($\varepsilon_1$) and imaginary part ($\varepsilon_2$) of the dielectric function of all the materials involved are shown in Fig. 2(c, f, i, l). The optical properties of NSTO are very similar to BTO film as they are both cubic perovskite materials with similar lattice parameters[22,34]. In previous studies, it has already been reported that Al impurity doped into ZnO films can act as effective n-type donors to generate free carriers[17]. By increasing the Al content, the carrier concentration in the AZO films increases, resulting in the decrease of the refractive index. In our results in Fig. 2(i, l), this decrease is observed as expected.

We found that the calculated Ψ and Δ are in good agreement with the experimental data; fits with mean-squared-error below 10 were achieved for NSTO, BTO and ZnO samples while for AZO film the mean-square error is less than 60 (Table I).

TABLE I: Thickness and surface roughness of NSTO, BTO, ZnO and AZO films extracted from SE analysis. Same information extracted by different characterization techniques (X-Ray reflectivity, profilometer, AFM) for comparison.

| Method / Film | Thickness (nm) | | | Roughness (nm) | | MSE |
|---|---|---|---|---|---|---|
| | XR Reflectivity | Profilometer | Spectroscopic Ellipsometry | Spectroscopic Ellipsometry | AFM | |
| 0.5 % Nb-doped SrTiO$_3$ | - | - | 1000 | - | - | 6.859 |
| BaTiO$_3$ (BTO) | 120 | - | 115.562 | - | 0.41 | 9.662 |
| ZnO | - | 300 | 301.540 | 6.019 | 3.87 | 9.188 |
| 4% Al-doped ZnO (AZO) | - | 340 | 333.448 | 6.741 | 3.93 | 59.71 |

For ZnO and AZO a three-layer model (surface roughness/material layer/MgO substrate) was used while for BTO a two-layer (material layer/MgO substrate) and for NSTO a one-layer model were used to extract the optical constants of the materials; the optical constants of MgO are well known[22,35]. The substrates NSTO and MgO were about 1mm thick and have been treated as semi-infinite.

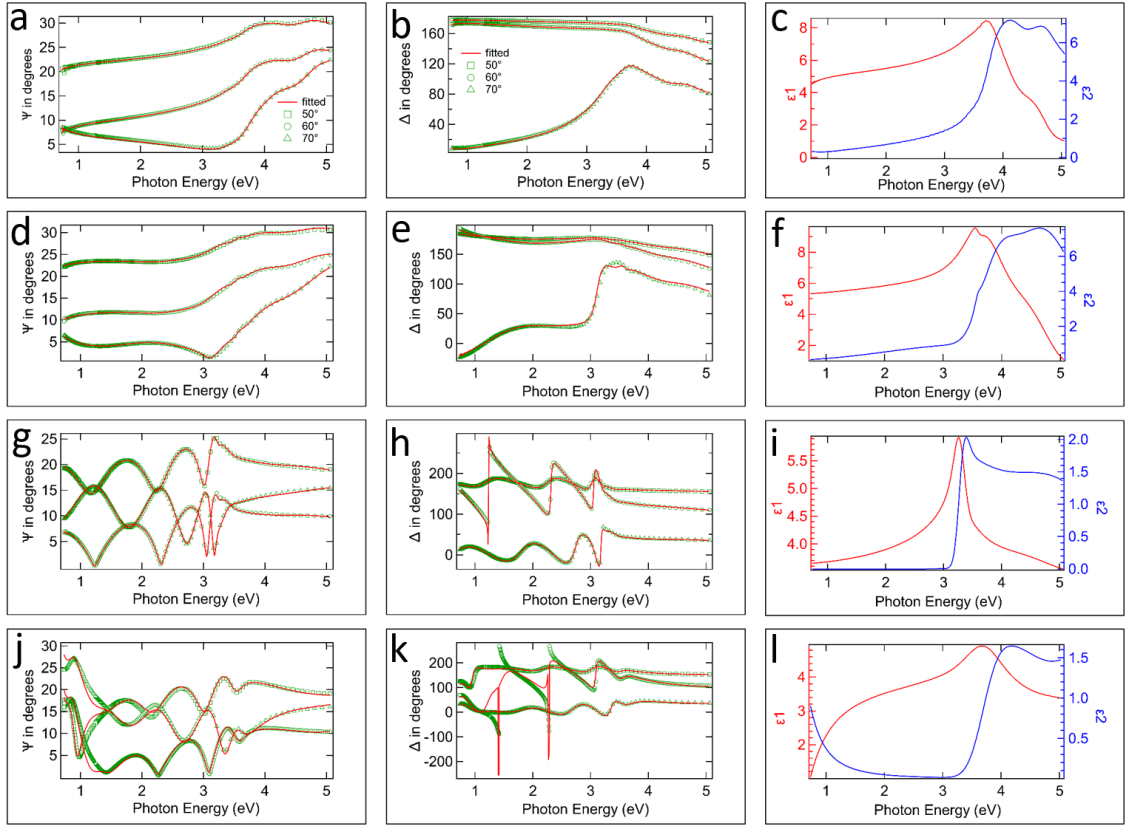

FIG. 2. Ψ, Δ ellipsometric angles and dielectric constants (ε1 and ε2) of the (a-c) conductive substrate NSTO, (d-f) insulating BTO film on NSTO, (g-i) ZnO film on MgO substrate and (j-l) 4% AZO on MgO substrate.

## B. Modeling of the response of AZO/BTO/NSTO multilayers

### 1. Bias-induced charge injection

After the optical characterization of each layer, we calculated the ellipsometric angles, Ψ and Δ, of the multilayer stack, using the model shown in Fig. 3. For the simulations, the thickness of AZO was set at 10 nm ($d_{AZO}$), the thickness of BTO at 100 nm ($d_{BTO}$) and its permittivity ($\varepsilon_r$) at 500.

The charge accumulation and as a sequence the change in carrier density were estimated from the capacitor equations. Then we assume the carrier density so-obtained (dependent only on the applied bias) as constant with respect to frequency and insert this in the model for the frequency dependent dielectric constant, used for the simulation of SE curves. Therefore, the modifications observed in SE are entirely due to the change in carrier density and therefore to the applied bias.

For the calculation of the carrier density of AZO, we used the formula of the Drude model (Eq. 2, 4). The dielectric constant of AZO ($\varepsilon = 4.07$) was calculated from the extracted $\varepsilon_1$ and $\varepsilon_2$ in the previous paragraph. For $E_{band\ gap}$ equal to 3 eV, $\varepsilon_1 = 4.069$ and $\varepsilon_2 = 0.015$.

$$\varepsilon_{n_{Drude}} = -\frac{ABr}{E^2 + iBrE} \Rightarrow \varepsilon^2 = \frac{A^2Br^2}{E^4 + Br^2E^2} \xrightarrow{A >> Br} \varepsilon^2 = \frac{A^2Br^2}{E^4} \Rightarrow \varepsilon = \frac{ABr}{E^2} \quad (2)$$

$$E = \hbar\omega = \hbar\sqrt{\frac{N_e q^2}{m^* \varepsilon_0}} \quad (3)$$

From Eq. 2 and 3 $\Rightarrow N_e = \frac{ABrm^*\varepsilon_0}{\varepsilon\hbar^2 q^2} \quad (4)$

In the case of the biased capacitor, the applied voltage was set in the range of 0-0.5 V. We have limited the study to a maximum of 0.5 V which is realistic for real devices and can be considered relatively safe for the dielectric barrier.

The variations on the surface charge density upon bias, $N_V$, were calculated by dividing the surface charge density, $\sigma_e$ (Eq. 5) by the thickness of AZO, $d_{AZO}$, as it is shown in Eq. 6:

$$\sigma_e = \frac{V_g \varepsilon_r \varepsilon_0}{d} \quad (5)$$

$$N_V = \frac{V_g \varepsilon_r \varepsilon_0}{dd_{AZO}} = \frac{\sigma_e}{d_{AZO}} \quad (6)$$

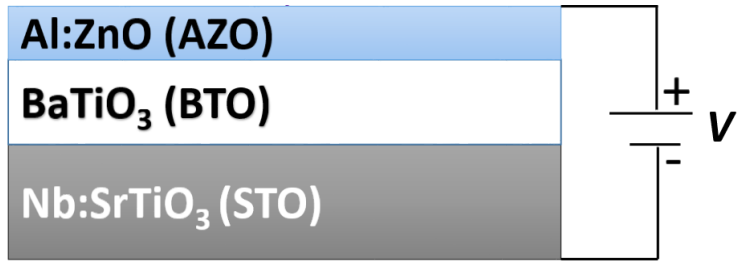

FIG. 3. Three-layer system used for spectroscopic ellipsometry modeling. The variations in the carrier density of AZO ($N_e$) upon dc gating (0.1-0.5V) were calculated.

## 2. Ellipsometry vs bias

The optical properties of each individual layer were combined in a model as it is shown in Fig. 3 for the extraction of the ellipsometric angles, Ψ and Δ, of the proposed capacitor (Fig. 4). The surface roughness factor is omitted from the model for the thin AZO film in order to simplify the treatment.

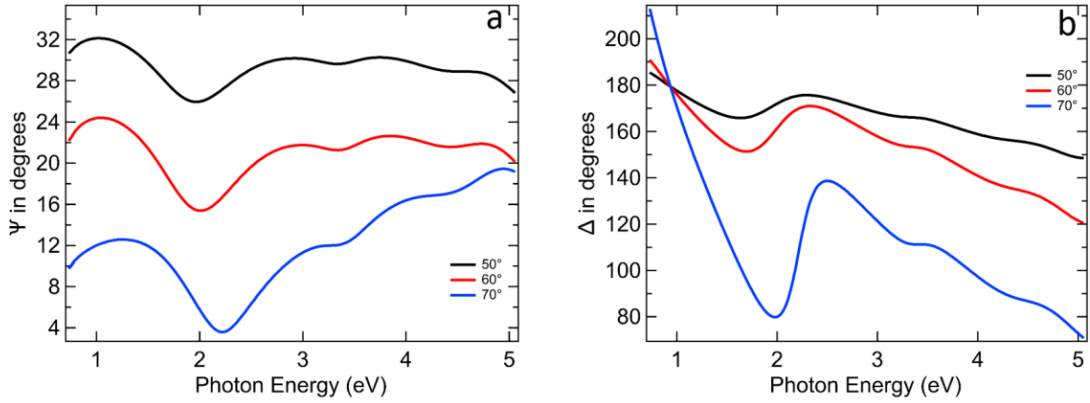

FIG. 4. The ellipsometric angles, (a) Ψ and (b) Δ, of the proposed capacitor (unbiased), as a function of the photon energy, extracted from the model shown in Fig. 2.

In Table II, the calculated variations of the carrier density upon bias are shown. As we increase the applied voltage, the surface charge density is also increased which results in higher carrier density variations. By applying voltages of ±0.1, ±0.3 and ±0.5 V, the variations are increased from 3.5 to 17.2 % compared to the carrier density, $N_e$, at zero voltage. The positive and negative values of the applied voltage are related to the two opposite directions of current, depending on the circuit connection.

TABLE II. Calculations of the variations on the carrier density ($N_V$) of the top gate of the capacitor (AZO), upon bias.

| Applied Voltage, V (V) | Surface charge density, $\sigma_e$ (C/cm²) | Carrier density variations, $N_v$ $N_v=\sigma_e/d_{AZO}$ (cm⁻³) | Variations, $N_v/N_e$ (%) |
|---|---|---|---|
| ± 0.1 | ± 0.28·10¹³ | ± 0.28·10¹⁹ | ±3.5 |
| ± 0.3 | ± 0.83·10¹³ | ± 0.83·10¹⁹ | ±10.4 |
| ± 0.5 | ± 1.38·10¹³ | ± 1.38·10¹⁹ | ±17.2 |

According to the Drude function of the modeling of 4% AZO, $A = 6.89 eV$ and $Br = 0.24 eV$, in the case of the unbiased capacitor. From the Eq. 4, $N_e$ is equal to $\sim 8.00 \cdot 10^{19} \, cm^{-3}$. For the simulations of the biased capacitor, the calculated values of $A$ that correspond to the $N_e$ values upon bias are shown in Table III.

TABLE III. Calculations of the density of carriers ($N_e$) and the parameter $A$ of the Drude formula, upon bias.

| Voltage applied (V) | Carrier density, $N_e$ (cm⁻³) | Parameter A (eV) |
|---|---|---|
| + 0.1 | 8.28·10¹⁹ | 7.13 |
| -0.1  | $7.72 \cdot 10^{19}$ | 6.68 |
| +0.3  | $8.83 \cdot 10^{19}$ | 7.60 |
| -0.3  | $7.17 \cdot 10^{19}$ | 6.18 |
| +0.5  | $9.38 \cdot 10^{19}$ | 8.08 |
| -0.5  | $6.62 \cdot 10^{19}$ | 5.70 |

Figure 5 shows the variations of $\Psi$ and $\Delta$ (at angles of incidence of 50°, 60° and 70°), upon gating in different voltages, as a function of photon energy. The variations are greater for angle of incidence of 70° and gating of 0.5V. It is observed that these variations are more pronounced in the near-infrared region (NIR), as expected considering the spectral dependence of the Drude component.

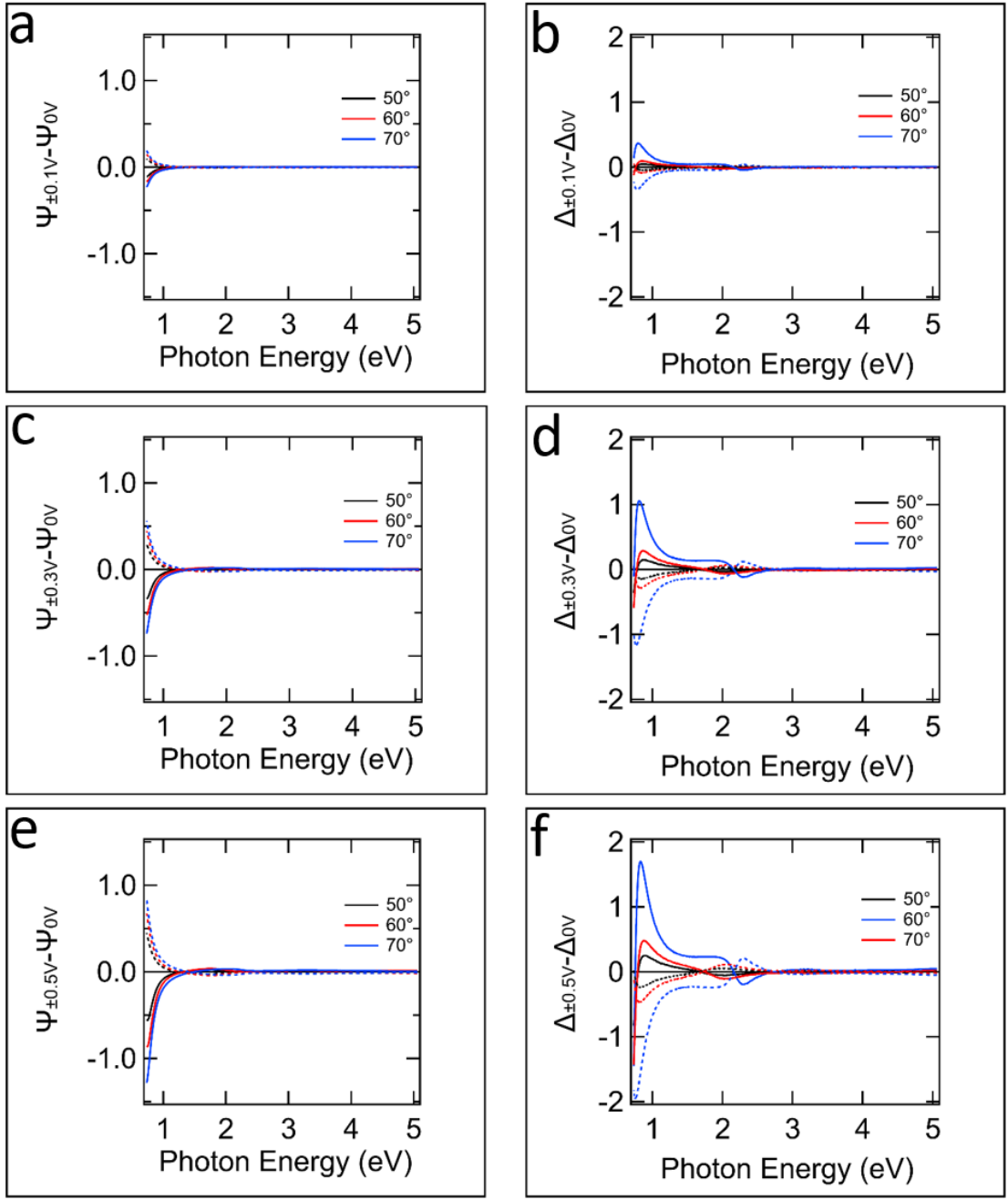

FIG. 5. Variations of the ellipsometric angles, (a) Ψ and (b) Δ, of the proposed capacitor, as a function of the photon energy, at angles of incidence of 50°, 60° and 70°, extracted from the model shown in Fig. 2. The solid lines represent the variations upon positive bias and the dashed lines the variations upon negative bias (black lines: ±0.1V, red lines: ±0.3V and blue lines: ±0.5V).

According to the results in Fig.5, Infrared SE is required for the detection of the variations in the dielectric constants of AZO upon bias. However, provided the model assumptions are correct, these effects are well within reach of SE.

## IV. SUMMARY AND CONCLUSIONS

In this work, SE was applied for the investigation of the optical properties of NSTO substrate, BTO film on NSTO, as well as, ZnO and 4% Al-doped ZnO films on MgO substrates. This study led to the modeling of a capacitor with BTO as the insulating layer, NSTO as the conductive bottom gate and 4%AZO film as the top gate. Simulations were performed in order to investigate the variations in the optical response of the proposed capacitor, upon voltage. These simulations could be a very useful tool for the determination of the optimum conditions to successfully monitor the variations in the optical properties of TCOs upon gating. The successful fabrication of the AZO/BTO/NSTO hybrid system with tunable properties upon gating is a first step for the fabrication of a prototype device for plasmo-electronics.

## ACKNOWLEDGMENTS

This project has received funding from the European Union's Horizon 2020 research and innovation programme under the Marie Skłodowska-Curie grant agreement N°799126. In addition, this work has been partly performed in the framework of the nanoscience foundry and fine analysis (NFFA-MIUR Italy) facility.